 \message
 {JNL.TEX version 0.92 as of 6/9/87.  Report bugs and problems to Doug Eardley.}
 
 \catcode`@=11
 \expandafter\ifx\csname inp@t\endcsname\relax\let\inp@t=\input
 \def\input#1 {\expandafter\ifx\csname #1IsLoaded\endcsname\relax
 \inp@t#1%
 \expandafter\def\csname #1IsLoaded\endcsname{(#1 was previously loaded)}
 \else\message{\csname #1IsLoaded\endcsname}\fi}\fi
 \catcode`@=12

 
 
 \font\twelverm=cmr10 scaled 1200    \font\twelvei=cmmi10 scaled 1200
 \font\twelvesy=cmsy10 scaled 1200   \font\twelveex=cmex10 scaled 1200
 \font\twelvebf=cmbx10 scaled 1200   \font\twelvesl=cmsl10 scaled 1200
 \font\twelvett=cmtt10 scaled 1200   \font\twelveit=cmti10 scaled 1200
 \font\twelvesc=cmcsc10 scaled 1200  \font\twelvesf=cmssdc10 scaled 1200
 
 
 \def\twelvepoint{\normalbaselineskip=12.4pt plus 0.1pt minus 0.1pt
   \abovedisplayskip 12.4pt plus 3pt minus 9pt
   \abovedisplayshortskip 0pt plus 3pt
   \belowdisplayshortskip 7.2pt plus 3pt minus 4pt
   \smallskipamount=3.6pt plus1.2pt minus1.2pt
   \medskipamount=7.2pt plus2.4pt minus2.4pt
   \bigskipamount=14.4pt plus4.8pt minus4.8pt
   \def\rm{\fam0\twelverm}          \def\it{\fam\itfam\twelveit}%
   \def\sl{\fam\slfam\twelvesl}     \def\bf{\fam\bffam\twelvebf}%
   \def\mit{\fam 1}                 \def\cal{\fam 2}%
   \def\sc{\twelvesc}               \def\tt{\twelvett}
   \def\sf{\twelvesf}
   \textfont0=\twelverm   \scriptfont0=\tenrm   \scriptscriptfont0=\sevenrm
   \textfont1=\twelvei    \scriptfont1=\teni    \scriptscriptfont1=\seveni
   \textfont2=\twelvesy   \scriptfont2=\tensy   \scriptscriptfont2=\sevensy
   \textfont3=\twelveex   \scriptfont3=\twelveex  \scriptscriptfont3=\twelveex
   \textfont\itfam=\twelveit
   \textfont\slfam=\twelvesl
   \textfont\bffam=\twelvebf \scriptfont\bffam=\tenbf
   \scriptscriptfont\bffam=\sevenbf
   \normalbaselines\rm}
 

 
 \def\beginlinemode{\endmode
   \begingroup\parskip=0pt \obeylines\def\\{\par}\def\endmode{\par\endgroup}}
 \def\beginparmode{\endmode
   \begingroup \def\endmode{\par\endgroup}}
 \let\endmode=\par
 {\obeylines\gdef\
 {}}
 \def\singlespace{\baselineskip=\normalbaselineskip}
 
 \def\oneandahalfspace{\baselineskip=\normalbaselineskip
   \multiply\baselineskip by 3 \divide\baselineskip by 2}
 \def\doublespace{\baselineskip=\normalbaselineskip \multiply\baselineskip by 2}

 \newcount\firstpageno
 \firstpageno=2
 \footline={\ifnum\pageno<\firstpageno{\hfil}\else{\hfil\twelverm\folio
  \hfil}\fi}
 \def\toppageno{\global\footline={\hfil}\global\headline
   ={\ifnum\pageno<\firstpageno{\hfil}\else{\hfil\twelverm\folio\hfil}
  \fi}}
 \let\rawfootnote=\footnote              
 \def\footnote#1#2{{\rm\singlespace\parindent=0pt\parskip=0pt
   \rawfootnote{#1}{#2\hfill\vrule height 0pt depth 6pt width 0pt}}}
 \def\raggedcenter{\leftskip=4em plus 12em \rightskip=\leftskip
   \parindent=0pt \parfillskip=0pt \spaceskip=.3333em \xspaceskip=.5em
   \pretolerance=9999 \tolerance=9999
   \hyphenpenalty=9999 \exhyphenpenalty=9999 }
 \def\dateline{\rightline{\ifcase\month\or
   January\or February\or March\or April\or May\or June\or
   July\or August\or September\or October\or November\or December\fi
   \space\number\year}}
 \def\received{\vskip 3pt plus 0.2fill
  \centerline{\sl (Received\space\ifcase\month\or
   January\or February\or March\or April\or May\or June\or
   July\or August\or September\or October\or November\or December\fi
   \qquad, \number\year)}}
 
 
 \hsize=6.5truein
 \hoffset=0pt
 \vsize=8.9truein
 \voffset=0pt
 \parskip=\medskipamount
 \def\\{\cr}
 \twelvepoint            
 \doublespace            
 \overfullrule=0pt       
 
 
 
 
 \def\title                      
   {\null\vskip 3pt plus 0.2fill
    \beginlinemode \doublespace \raggedcenter \bf}
 
 \def\author                     
   {\vskip 3pt plus 0.2fill \beginlinemode
    \singlespace \raggedcenter\sl}    
 
 \def\affil                      
   {\vskip 3pt plus 0.1fill \beginlinemode
    \oneandahalfspace \raggedcenter \sl}
 
 \def\abstract                   
   {\vskip 3pt plus 0.3fill \beginparmode
    \oneandahalfspace ABSTRACT: }
 
 \def\endtitlepage               
   {\endpage                     
    \body}
 \let\endtopmatter=\endtitlepage
 
 \def\body                       
   {\beginparmode}               
 
 \def\head#1{                    
   \goodbreak\vskip 0.5truein    
   {\immediate\write16{#1}
    \raggedcenter \uppercase{#1}\par}
    \nobreak\vskip 0.25truein\nobreak}

 \def\beginitems{
 \par\medskip\bgroup\def\i##1 {\item{##1}}\def\ii##1 {\itemitem{##1}}
 \leftskip=36pt\parskip=0pt}
 \def\enditems{\par\egroup}
 
 \def\beneathrel#1\under#2{\mathrel{\mathop{#2}\limits_{#1}}}

 \def\refto#1{[#1]}
 
 \def\references                 
   {\head{References}            
    \beginparmode
    \frenchspacing \parindent=0pt \leftskip=1truecm
    \parskip=8pt plus 3pt \everypar{\hangindent=\parindent}}

 \gdef\refis#1{\item{#1.\ }}                     

 \gdef\journal#1, #2, #3, 1#4#5#6{           
 {\sl #1~}{\bf #2}, #3 (1#4#5#6)}            


 \def\endreferences{\body}
 
 \def\figurecaptions             
   {\endpage
    \beginparmode
    \head{Figure Captions}
 }

 \def\endpage                    
   {\vfill\eject}
 
 \def\endpaper                   
   {\endmode\vfill\supereject}


 \def\heading                            
   {\vskip 0.5truein plus 0.1truein      
    \beginparmode \def\\{\par} \parskip=0pt \singlespace \raggedcenter}

 \def\subheading                         
   {\vskip 0.25truein plus 0.1truein     
    \beginlinemode \singlespace \parskip=0pt \def\\{\par}\raggedcenter}

 \def\tag#1$${\eqno(#1)$$}
 
 \def\align#1$${\eqalign{#1}$$}

 \def\aligntag#1$${\gdef\tag##1\\{&(##1)\cr}\eqalignno{#1\\}$$
   \gdef\tag##1$${\eqno(##1)$$}}
 
 \def\endaligntag{}

 \def\overset #1\to#2{{\mathop{#2}\limits^{#1}}}
 \def\underset#1\to#2{{\let\next=#1\mathpalette\undersetpalette#2}}
 \def\undersetpalette#1#2{\vtop{\baselineskip0pt
 \ialign{$\mathsurround=0pt #1\hfil##\hfil$\crcr#2\crcr\next\crcr}}}

 
 \def\ref#1{Ref.~#1}                     
 \def\Ref#1{Ref.~#1}                     
 \def\[#1]{[\cite{#1}]}
 \def\cite#1{{#1}}
 \def\Eq#1{Eq.~\(#1)}                    
 \let\eq=\Eq                
 \def\(#1){(\call{#1})}
 \def\call#1{{#1}}
 \def\taghead#1{}
 \def\frac#1#2{{#1 \over #2}}

 \def\12{{1\over2}}

 \def\ie{{\it i.e.,\ }}

 \def\sla{\raise.15ex\hbox{$/$}\kern-.57em}
 \def\leaderfill{\leaders\hbox to 1em{\hss.\hss}\hfill}
 \def\twiddle{\lower.9ex\rlap{$\kern-.1em\scriptstyle\sim$}}
 \def\bigtwiddle{\lower1.ex\rlap{$\sim$}}
 \def\gtwid{\mathrel{\raise.3ex\hbox{$>$\kern-.75em\lower1ex\hbox{$\sim$}}}}
 \def\ltwid{\mathrel{\raise.3ex\hbox{$<$\kern-.75em\lower1ex\hbox{$\sim$}}}}
 \def\square{\kern1pt\vbox{\hrule height 1.2pt\hbox{\vrule width 1.2pt\hskip 3pt
    \vbox{\vskip 6pt}\hskip 3pt\vrule width 0.6pt}\hrule height 0.6pt}\kern1pt}
 \def\tdot#1{\mathord{\mathop{#1}\limits^{\kern2pt\ldots}}}
 
 \def\pmb#1{\setbox0=\hbox{#1}%
   \kern-.025em\copy0\kern-\wd0
   \kern  .05em\copy0\kern-\wd0
   \kern-.025em\raise.0433em\box0 }

 \catcode`@=11
 \newcount\r@fcount \r@fcount=0
 \newcount\r@fcurr
 \immediate\newwrite\reffile
 \newif\ifr@ffile\r@ffilefalse
 \def\w@rnwrite#1{\ifr@ffile\immediate\write\reffile{#1}\fi\message{#1}}
 
 \def\writer@f#1>>{}
 \def\referencefile{
   \r@ffiletrue\immediate\openout\reffile=\jobname.ref%
   \def\writer@f##1>>{\ifr@ffile\immediate\write\reffile%
     {\noexpand\refis{##1} = \csname r@fnum##1\endcsname = %
      \expandafter\expandafter\expandafter\strip@t\expandafter%
      \meaning\csname r@ftext\csname r@fnum##1\endcsname\endcsname}\fi}%
   \def\strip@t##1>>{}}

 \def\citeall#1{\xdef#1##1{#1{\noexpand\cite{##1}}}}
 \def\cite#1{\each@rg\citer@nge{#1}}     
 
 \def\each@rg#1#2{{\let\thecsname=#1\expandafter\first@rg#2,\end,}}
 \def\first@rg#1,{\thecsname{#1}\apply@rg}       
 \def\apply@rg#1,{\ifx\end#1\let\next=\relax
 \else,\thecsname{#1}\let\next=\apply@rg\fi\next}
 
 \def\citer@nge#1{\citedor@nge#1-\end-}  
 \def\citer@ngeat#1\end-{#1}
 \def\citedor@nge#1-#2-{\ifx\end#2\r@featspace#1 
   \else\citel@@p{#1}{#2}\citer@ngeat\fi}        
 \def\citel@@p#1#2{\ifnum#1>#2{\errmessage{Reference range #1-#2\space is bad}.%
     \errhelp{If you cite a series of references by the notation M-N, then M and
     N must be integers, and N must be greater than or equal to M.}}\else
  {\count0=#1\count1=#2\advance\count1 by1\relax\expandafter\r@fcite\the\count0,%
   \loop\advance\count0 by1\relax
     \ifnum\count0<\count1,\expandafter\r@fcite\the\count0,%
   \repeat}\fi}
 
 \def\r@featspace#1#2 {\r@fcite#1#2,}    
 \def\r@fcite#1,{\ifuncit@d{#1}
     \newr@f{#1}%
     \expandafter\gdef\csname r@ftext\number\r@fcount\endcsname%
                      {\message{Reference #1 to be supplied.}%
                       \writer@f#1>>#1 to be supplied.\par}%
  \fi%
  \csname r@fnum#1\endcsname}
 \def\ifuncit@d#1{\expandafter\ifx\csname r@fnum#1\endcsname\relax}%
 \def\newr@f#1{\global\advance\r@fcount by1%
     \expandafter\xdef\csname r@fnum#1\endcsname{\number\r@fcount}}
 
 \let\r@fis=\refis                       
 \def\refis#1#2#3\par{\ifuncit@d{#1}
 blank
    \newr@f{#1}%
    \w@rnwrite{Reference #1=\number\r@fcount\space is not cited up to now.}\fi%
   \expandafter\gdef\csname r@ftext\csname r@fnum#1\endcsname\endcsname%
   {\writer@f#1>>#2#3\par}}
 
 \def\ignoreuncited{
    \def\refis##1##2##3\par{\ifuncit@d{##1}%
      \else\expandafter\gdef\csname r@ftext\csname r@fnum##1\endcsname\endcsname%
      {\writer@f##1>>##2##3\par}\fi}}
 
 \def\r@ferr{\endreferences\errmessage{I was expecting to see
 \noexpand\endreferences before now;  I have inserted it here.}}
 \let\r@ferences=\references
 \def\references{\r@ferences\def\endmode{\r@ferr\par\endgroup}}
 
 \let\endr@ferences=\endreferences
 \def\endreferences{\r@fcurr=0
   {\loop\ifnum\r@fcurr<\r@fcount
     \advance\r@fcurr by 1\relax\expandafter\r@fis\expandafter{\number\r@fcurr}%
     \csname r@ftext\number\r@fcurr\endcsname%
   \repeat}\gdef\r@ferr{}\endr@ferences}
 
 
 \let\r@fend=\endpaper\gdef\endpaper{\ifr@ffile
 \immediate\write16{Cross References written on []\jobname.REF.}\fi\r@fend}
 
 \catcode`@=12

 \citeall\refto          
 \citeall\ref            %
 \citeall\Ref            %

\catcode`@=11
\newcount\tagnumber\tagnumber=0

\immediate\newwrite\eqnfile
\newif\if@qnfile\@qnfilefalse
\def\write@qn#1{}
\def\writenew@qn#1{}
\def\w@rnwrite#1{\write@qn{#1}\message{#1}}
\def\@rrwrite#1{\write@qn{#1}\errmessage{#1}}

\def\taghead#1{\gdef\t@ghead{#1}\global\tagnumber=0}
\def\t@ghead{}

\expandafter\def\csname @qnnum-3\endcsname
  {{\t@ghead\advance\tagnumber by -3\relax\number\tagnumber}}
\expandafter\def\csname @qnnum-2\endcsname
  {{\t@ghead\advance\tagnumber by -2\relax\number\tagnumber}}
\expandafter\def\csname @qnnum-1\endcsname
  {{\t@ghead\advance\tagnumber by -1\relax\number\tagnumber}}
\expandafter\def\csname @qnnum0\endcsname
  {\t@ghead\number\tagnumber}
\expandafter\def\csname @qnnum+1\endcsname
  {{\t@ghead\advance\tagnumber by 1\relax\number\tagnumber}}
\expandafter\def\csname @qnnum+2\endcsname
  {{\t@ghead\advance\tagnumber by 2\relax\number\tagnumber}}
\expandafter\def\csname @qnnum+3\endcsname
  {{\t@ghead\advance\tagnumber by 3\relax\number\tagnumber}}

\def\equationfile{%
  \@qnfiletrue\immediate\openout\eqnfile=\jobname.eqn%
  \def\write@qn##1{\if@qnfile\immediate\write\eqnfile{##1}\fi}
  \def\writenew@qn##1{\if@qnfile\immediate\write\eqnfile
    {\noexpand\tag{##1} = (\t@ghead\number\tagnumber)}\fi}
}

\def\callall#1{\xdef#1##1{#1{\noexpand\call{##1}}}}
\def\call#1{\each@rg\callr@nge{#1}}

\def\each@rg#1#2{{\let\thecsname=#1\expandafter\first@rg#2,\end,}}
\def\first@rg#1,{\thecsname{#1}\apply@rg}
\def\apply@rg#1,{\ifx\end#1\let\next=\relax%
\else,\thecsname{#1}\let\next=\apply@rg\fi\next}

\def\callr@nge#1{\calldor@nge#1-\end-}
\def\callr@ngeat#1\end-{#1}
\def\calldor@nge#1-#2-{\ifx\end#2\@qneatspace#1 %
  \else\calll@@p{#1}{#2}\callr@ngeat\fi}
\def\calll@@p#1#2{\ifnum#1>#2{\@rrwrite{Equation range #1-#2\space is bad.}
\errhelp{If you call a series of equations by the notation M-N, then M and
N must be integers, and N must be greater than or equal to M.}}\else%
 {\count0=#1\count1=#2\advance\count1 by1\relax\expandafter\@qncall\the\count0,%
  \loop\advance\count0 by1\relax%
    \ifnum\count0<\count1,\expandafter\@qncall\the\count0,%
  \repeat}\fi}

\def\@qneatspace#1#2 {\@qncall#1#2,}
\def\@qncall#1,{\ifunc@lled{#1}{\def\next{#1}\ifx\next\empty\else
  \w@rnwrite{Equation number \noexpand\(>>#1<<) has not been defined yet.}
  >>#1<<\fi}\else\csname @qnnum#1\endcsname\fi}

\let\eqnono=\eqno
\def\eqno(#1){\tag#1}
\def\tag#1$${\eqnono(\displayt@g#1 )$$}

\def\aligntag#1\endaligntag
  $${\gdef\tag##1\\{&(##1 )\cr}\eqalignno{#1\\}$$
  \gdef\tag##1$${\eqnono(\displayt@g##1 )$$}}

\def\eqalignno#1{\displ@y \tabskip\centering
  \halign to\displaywidth{\hfil$\displaystyle{##}$\tabskip\z@skip
    &$\displaystyle{{}##}$\hfil\tabskip\centering
    &\llap{$\displayt@gpar##$}\tabskip\z@skip\crcr
    #1\crcr}}

\def\displayt@gpar(#1){(\displayt@g#1 )}

\def\displayt@g#1 {\rm\ifunc@lled{#1}\global\advance\tagnumber by1
        {\def\next{#1}\ifx\next\empty\else\expandafter
        \xdef\csname @qnnum#1\endcsname{\t@ghead\number\tagnumber}\fi}%
  \writenew@qn{#1}\t@ghead\number\tagnumber\else
        {\edef\next{\t@ghead\number\tagnumber}%
        \expandafter\ifx\csname @qnnum#1\endcsname\next\else
        \w@rnwrite{Equation \noexpand\tag{#1} is a duplicate number.}\fi}%
  \csname @qnnum#1\endcsname\fi}

\def\ifunc@lled#1{\expandafter\ifx\csname @qnnum#1\endcsname\relax}

\let\@qnend=\end\gdef\end{\if@qnfile
\immediate\write16{Equation numbers written on []\jobname.EQN.}\fi\@qnend}

\catcode`@=12



\newcount\notenumber
\def\clearnotenumber{\notenumber=0}
\def\note{\advance\notenumber by1 \footnote{$^{\the\notenumber}$}}
\clearnotenumber

\def\chap{\mathaccent 94}

\def\til{\mathaccent "7E }

\def\lim#1 {\displaystyle{\mathop{\ell im}_{#1}}}

\def\bra{\langle}
\def\ket{\rangle}
\def\gcall#1{(\call{#1})}
\def\dt{ {\partial \over {\partial t} } }
\def\dR{ {\partial \over {\partial {\bf R} } } }
\def\dV{ {\partial \over {\partial {\bf V} } } }
\def\dr#1{ {\partial \over {\partial {\bf r_{#1}} } } }
\def\dU{ {\partial \over {\partial {\bf U} } } }

\def\bR{ {\bf R} }
\def\br#1{ {\bf r_{#1}} }
\def\bV{ {\bf V} }
\def\bv#1{ {\bf v_{#1}}}
\def\dtau{ {\partial \over {\partial \tau} } }
\def\dta#1{ {\partial \over {\partial \tau_{#1}} } }
\def\dX{ {\partial \over {\partial {\bf X} } } }
\def\dx#1{ {\partial \over {\partial {\bf x_{#1}} } } }
\def\bX{ {\bf X} }
\def\bx#1{ {\bf x_{#1}} }
\def\bU{ {\bf U} }
\def\bu#1{ {\bf u_{#1}}}

\def\sig1{ {\chap{\bf \sigma} } }
\def\sec{ \xi^2 }
\def\dia{ \xi }
\def\Tm{ \overline {T}_{-} }
\def\Tb{ \overline {T}_{-} }
\def\Tme { \overline {T}_{-}^{\epsilon} }

\def\Fpm {{\bf\cal F}_{\pm}}
\def\Fmpn {{\bf\cal F}_{\mp}}
\def\eq{ {(eq\vert\bX)}}
\def\Eq{ {(eq\vert\bR)}}
\def\Fmn {{\bf\cal F}_{-}}
\def\Fpn {{\bf\cal F}_{+}}
\def\Fmoy { \overline \Fpn}
\def\Ftot { \overline {\bf\cal F}_f}
\def\Lf { {\cal L}_f}
\def\LB { {\cal L}_B}
\def\Cdyn { {\cal C}^{dyn} }
\def\Vmoy { \overline{\bV}_B }
\def\tVmoy { \til{\overline{\bV}}_B }
\def\Vchap { \chap{\bV}}

\null
\vfill
\oneandahalfspace
\head{\bf Microscopic Derivation of Non-Markovian Thermalization of a Brownian Particle}

\vskip 15pt
\author Lyd\'eric Bocquet and Jaros\l aw Piasecki $\dag$  

\affil{ Laboratoire de Physique, Ecole Normale Sup\'erieure
de Lyon (URA CNRS 1325), 
46 All\'ee d'Italie, 69007  Lyon (France)}

\affil{ $\dag$ Institute of Theoretical Physics, Warsaw University,
Hoza 69, 00-681 Warsaw (Poland)}

\abstract

In this paper, the first microscopic approach to the Brownian
motion is developed in the case where the mass density of the suspending
bath is of the same order of magnitude as that of the Brownian (B) particle. Starting
from an extended Boltzmann equation, which describes correctly the
interaction with the fluid, we derive
systematicaly via the multiple time-scale analysis a reduced
equation controlling the thermalization of the B particle, \ie the 
relaxation towards the Maxwell distribution in velocity space. In
contradistinction to the Fokker-Planck equation, the derived new evolution
equation is non-local
both in time and in velocity space, owing to correlated recollision events 
between the fluid and particle B. 
In the long-time limit, it describes a non-markovian generalized
Ornstein-Uhlenbeck process. However, in spite of this
complex dynamical behaviour, the Stokes-Einstein law relating the
friction and diffusion coefficients is shown to remain valid. A microscopic
expression for the friction coefficient is derived, which
acquires the form of the Stokes law in the limit where the mean-free path in the
gas is small compared to the radius of particle B.

submitted to Journal of Statistical Physics 

\dateline

\vfill
\eject

\smallskip 
\endtopmatter
{\bf 1. Introduction}

This paper is concerned with a microscopic theory of
the Brownian motion performed by a massive particle suspended in
a gas of much lighter particles. 
Because
of the mass and length-scale difference between the two
components, one expects to eliminate the fluid variables from 
the description of the system and obtain a closed equation
for the B particle \refto{Wax, Einstein, VanK}.
Traditionally, this task is done by assuming
that the dynamical properties of the Brownian
particle (B) evolve on a time-scale 
much longer than the characteristic time-scale of the fluid.
This approach leads to the well-known
Fokker-Planck equation, which governs the time evolution of
the distribution function $f(\bR,\bV;t)$ of the position $\bR$ and velocity
$\bV$ of particle B :
$$
\left( \dt + \bV \cdot \dR\right) f (\bR,\bV;t)=
\zeta \ \dV \cdot \left( \bV + {{k_BT}\over M} \dV \right) f(\bR,\bV;t)
\eqno(FP)
$$
$M$ denotes the mass. In this equation, the fluid
enters only through the friction coefficient $\zeta$ and the temperature $T$. 

An
equivalent description can be obtained starting from the
stochastic Langevin equation
$$
M {d \bV \over {d t} } = -M \zeta \bV (t) + \til{\bf F} (t)
\eqno(Langevin)
$$
where $\bV (t)$ is the velocity of B and $\til{\bf F} (t)$ is the
fluctuating part of the force exerted by the fluid, assumed to have
a white spectrum. This equation leads to an exponential decay of
the velocity autocorrelation function with a relaxation time given
by $\zeta^{-1}$ \refto{Wax, Einstein}. 

On the other hand, some attempts at a fully microscopic description
of the dynamics have been made \refto{Lebowitz, Cukier, Mercer, Deutch,
Papiers_I,Autres}. It has indeed been possible to get rid of the stochastic
assumption and obtain the Fokker-Planck equation by a systematic
expansion of the dynamics of the complete system, bath + B particle,
in powers of the square root of the mass ratio $m/M$, where $m$ is the fluid particle mass.
Such derivations lead in particular to a microscopic formula for the
friction coefficient in terms of the autocorrelation function of the
force acting on the B particle.

However, as pointed out by many authors \refto{Hauge, Hinch, Masters,
Roux}, but already
by Lorentz in 1911 \refto{Lorentz}, the description based on eqs. \gcall{FP} and \gcall{Langevin} can only be valid if
the ratio $\rho/\rho_B$ of the mass density of the fluid, $\rho$, and that of the B 
particle, $\rho_B$, is very small. Indeed, as 
noticed above, the velocity of particle B relaxes on a time-scale 
$\tau_V\sim 
\zeta^{-1}$. If Stokes' law is assumed, this leads to
$\tau_V\sim M/\eta \Sigma$, where $\eta$ is the viscosity of the fluid,
and $\Sigma$ the B particle diameter. On the other hand, a typical
hydrodynamic time of the fluid is of the order $\tau_f \sim \Sigma ^2/(\eta/\rho)$,
which is the time for a shear flow to propagate over the distance $\Sigma$.
These rough estimates give $\tau_f/\tau_V\sim \rho/\rho_B$, so that
the assumption of a wide time-scale separation is only justified if
this ratio is small. This condition is unfortunately 
far from the
experimental situation, where the mass density ratio $\rho/\rho_B$ is
taken rather close to unity to avoid sedimentation of particle B.
In this case, the fluid dynamics must contain slowly decaying modes, which
relax on the same time scale as the velocity of the B particle.
In other words, non-Markovian effects are expected and the validity of the 
Fokker-Planck or Langevin equations becomes doubtful \note{Let us note
however that the Smoluchowski equation, describing the spatial evolution
of the B particle is not affected by these arguments, since the position
of the B particle relaxes on a time-scale much longer than the velocity
(or fluid) relaxation times.} \refto{Masters}.

Several attempts have been made to overcome these difficulties and determine 
the dynamical evolution in the general case, that is 
for any (finite) $\rho/\rho_B$ \refto{Hauge, Hinch, Bedeaux}.
The main idea underlying all these works is that the slowly decaying fluid modes
result from the momentum conservation law for
the fluid particles. 
A correct description should therefore treat both fluid and B particle
variables on the same level.
This can be done for example
by using
fluctuating hydrodynamics for the fluid motion, with appropriate boundary conditions on the surface of the suspended
B particle \refto{Hauge}. These approaches 
lead to a non-Markovian Langevin equation, involving
the time-dependent friction coefficient $\zeta(t)$ :
$$
M {d \bV \over {d t} } = -M \int_0^t d\tau\ \zeta(t-\tau)\ \bV (t) + \til{\bf F} (t)
\eqno(NMLangevin)
$$
\vskip 2mm
We present here the first systematic, microscopic approach which treats the B particle and the
fluid at the same level, clarifying and providing
the fundamental basis for such phenomenological descriptions.
The problem is reconsidered within the kinetic
theory of gases. The system consists of a single, large, heavy
hard sphere (mass $M$, diameter $\Sigma$) suspended in a fluid of small, light hard spheres (mass $m$, diameter $\sigma$). The dynamics of the system
will be assumed to be governed by an extended Boltzmann equation,
which correctly takes the gas-B particle collisions into account. The same type of equation
was used by van Beijeren and Dorfman to develop
the kinetic theory of hydrodynamic flows and in particular to clarify the
dynamical origin of the hydrodynamic Stokes law for the friction 
coefficient \refto{DVB}. The extended Boltzmann equation is expected to be correct in the
Grad limit, defined for the gas with number density $n$ by $n\rightarrow \infty$,
$\sigma\rightarrow 0$, the mean free path  $\ell=(n\sigma^2)^{-1}=constant$ 
\refto{Spohn}. We choose the 
diameter $\Sigma$ of the B particle to fix the length scale. The condition
of a constant mean free path thus imposes
$$
{\Sigma \over \ell} = n\sigma^2 \Sigma =const
\eqno(Cond1)
$$
Moreover, and unlike all previous work, we require from the begining the condition
of essentially equal mass density of the gas and the B particle. In other words, the ratio $\rho/\rho_B$ will be kept finite even in
the small mass ratio limit, which amounts to assume
$$
{M \over \Sigma^3} = const \times m n
\eqno(Cond2)
$$
Finally, we define the small parameter corresponding to the Brownian
limit
$$
\epsilon =\left( {m \over M} \right)^{1/2} \ll 1
\eqno(Massratio)
$$

Introducing the diameter ratio $\kappa=\sigma/\Sigma$, we characterize the
asymptotic regime \gcall{Cond1}-\gcall{Massratio} by the limit :
$$
\cases{
\epsilon \rightarrow 0 & \cr
\kappa \sim \epsilon, \ \ n\Sigma^3 \sim \epsilon^{-2} & \cr
}
\eqno(Deflimit)
$$
To simplify the calculations and without loss of generality, we will assume in the following that $\epsilon=\kappa$ and $n\Sigma^3 = \epsilon^{-2}$. This can be looked upon as redefining
the parameters of the system ($\Sigma$, $M$, $n$, ...), to set the two 
constants introduced in eqs. \gcall{Cond1} and \gcall{Cond2} equal to one. 
The correct scaling will however be recovered in the final results.
 
Starting from the extended Boltzmann equation, we will perform a systematic
expansion of the dynamical properties of the system in powers of
$\epsilon$, in order to obtain a closed equation for the distribution
function of particle B. 

The following results will be obtained :

(i) On the shortest time-scale, $t \sim \tau_V \sim \tau_f$, the B particle ``does not move'' (in position space), whereas its velocity distribution
relaxes in a thermalization process.

(ii) The equation (65) characterizing the relaxation of the velocity distribution
is not a Fokker-Planck equation. It exhibits a complex non-markovian
character, which stems from building up the dynamical friction force 
by recollision events between the gas and the 
B particle.

(iii) We will be able to express the autocorrelation function (ACF) of the velocity of the B particle, in terms of the time-dependent friction
coefficient. A microscopic formula for this coefficient follows from our
analysis. The final result justifies the phenomenological expression
for the velocity ACF,
based on fluctuating hydrodynamics. 

(iv) In spite of non-markovian effects, the Stokes-Einstein relation, expressing the diffusion constant of
the B particle in terms of the friction coefficient, is explicitly
derived.

\vskip 0.4in{\bf 2. Kinetic Equations}

In the study of the dynamics of the B particle immersed in a bath of $N$ small spheres, the short-hand notation will be used :
$$
B \equiv (\bR,\bV); \ \ \ i\equiv (\br{i},\bv{i}), \ \ i=1,2,\dots,N
\eqno(Defnot)
$$
for the positions and velocities of the B particle and the N gas particles. 

The temperature $T$ of the gas introduces characteristic measures,
$\sqrt{k_BT/M}$ and $\sqrt{k_BT/m}$, of the 
velocities of the B and fluid particles, respectively. On the other hand, the characteristic length scale
in the system is given by the diameter $\Sigma$ of the B particle. In order
to study the asymptotic regime \gcall{Deflimit}, it is convenient to use dimensionless variables defined by
$$
 \vbox{ 
   \halign{
      \tabskip=8mm # \hfil & # \hfil \cr
        
      $\bV = \sqrt{ {{k_BT} \over M} }\ \bU$, & $\bv{j}= \sqrt{ {{k_BT} \over m} }\ \bu{j}$  \cr
      & \cr
      $\bR = \Sigma \bX$, & $\br{j} = \Sigma \bx{j}$ \cr
      & \cr
   }
 }
\eqno(Var)
$$
The shortest time-scale in the system is fixed by the gas-B particle
collision frequency, and we introduce accordingly a dimensionless time variable
$\tau$ by
$$
t = {1 \over {n\Sigma^2}} \sqrt{ {m\over k_BT} }\ \tau
\eqno(Tau)
$$ 

Our aim is to determine the dynamical evolution
of the B particle distribution function $f_B(B;t)$. To this end, we
need to introduce the joint distribution, $f_{B1}(B,1;t)$, representing at time $t$ the number
density of fluid particles in state $1$, when at the same time particle B
is in state $B$. Finally, we define the conditional distribution function
of the gas, $f_{1\vert B}$, by :
$$
f_{B1}(B,1;t)=f_{B}(B;t)\ f_{1\vert B}(1 \vert B;t)
\eqno(Defcond)
$$

Under the dimensional scaling \gcall{Var}, the dimensionless distribution functions, $F_B$ and $\til{F}_{1\vert B}$, are defined by the relations
$$
F_B (B,\tau) = \Sigma^3 (k_BT/M)^{3/2} f_B (B;t) 
\eqno(FBdim)
$$
and 
$$
\til{F}_{1\vert B} (1\vert B;\tau) = \Sigma^3 (k_BT/m)^{3/2} f_{1\vert B} (1\vert B;t) 
\eqno(Fctil)
$$
However, whereas $F_B(B;\tau)$ is essentially of order unity, $\til{F}_{1\vert B}$
does diverge as $\epsilon \rightarrow 0$. Indeed, far from the suspended
sphere B, the conditional distribution function $f_{1\vert B}$ is of the order of the
density $n$ of the gas, which implies (according to condition \gcall{Deflimit}) :
$$
\til{F}_{1\vert B} \sim n\Sigma ^3 \sim \epsilon^{-2}
\eqno(Fctilorder)
$$
Having in view a perturbation expansion, we thus introduce rescaled distribution functions $F_{1\vert B}(1\vert B;\tau)$ and $F_{B1}(B,1;\tau)$ defined as
$$
\eqalign{
&F_{1\vert B}(1 \vert B;\tau) = \epsilon ^2 \ \til{F}_{1\vert B} (1\vert B;\tau)\cr
&F_{B1}(B,1;\tau) =\epsilon ^2 \  F_B(B;\tau) \til{F}_{1\vert B}(1 \vert B;\tau) \cr
}
\eqno(DefFc)
$$
which are of order unity when the small parameter $\epsilon$ goes to
zero.

The scaled distribution functions evolve in time according to the 
coupled equations :
$$
\left( \dtau + \epsilon^{3} \bU \cdot \dX \right ) F_B (B;\tau)
=\int d{\bf 1} \ \Tme (B,1) F_{B1} (B,{\bf 1};\tau),
\eqno (EquB)
$$
$$
\eqalign{
\biggl\{ \dtau + \epsilon^3 \bU \cdot \dX &+ \epsilon^2 \left(\bu1\cdot \dx{1} - \Tme (B,1)\right) \biggr\}
F_{B1} (B,1;\tau) =\cr
&=\int d{\bf 2}\ \left\{ \Tme(B,2)+\epsilon^2 \Tb(1,2)\right\} F_{B12} (B,1,2;\tau)
}
\eqno(EquB1)
$$
where the factorization of the three-particle distribution 
$$
F_{B12} (B,1,2;t)\simeq F_B(B;t) F_{1\vert B}(1\vert B;t) F_{1\vert B}(2\vert B;t)
\eqno(Fact)
$$
is expected to hold in the Grad limit (see the discussion leading to \gcall{Cond1}).

Equation \gcall{EquB} expresses the fact that the B particle
state changes in
time through free-motion (left-hand side (l.h.s) of the equation) and
through collisions with gas particles (right-hand side (r.h.s) of the 
equation). The second equation \gcall{EquB1}
states that the two-particle distribution, $F_{B1}$, evolves
in time, owing to free motion and collisions between
the pair $(B,1)$ (l.h.s of the equation), and also owing to collisions of this
pair with the surrounding gas particles (r.h.s of the equation).

These evolution equations involve two hard-sphere 
collision operators, $\Tme (B,1)$ and $\Tb (1,2)$. The first of them characterises the
effect of binary collisions between the B particle and a gas
particle. It reads :
$$
\eqalign {
\Tme (B,1) = & \sec
\int d\sig1 \ \left[ \left( \epsilon \bU - \bu1\right) \cdot \sig1 \right]
\theta \left[ \left( \epsilon\bU - \bu1\right) \cdot \sig1 \right] \cr
&\times \left\{ \delta\left(\bX-\xi \sig1
-\bx1\right) b_{\sig1}^{\epsilon} (B,1) -
\delta\left(\bX+\xi \sig1
-\bx1\right) \right\}
}
\eqno (DefTme)
$$
where $\xi=1/2$ is the dimensionless radius of the B particle; the operator $b_{\sig1}^{\epsilon}$ transforms
the pre-collisional velocities into post-collisional
velocities :
$$
\left[b_{\sig1}^{\epsilon} (B,1) \chi \right] (\bU,\bu1) =
\chi \left( \bU -{{2\epsilon} \over {1+\epsilon^2}}\left[ \left(
\epsilon \bU-\bu1\right)\cdot \sig1 \right] \sig1, \bu1 + {2\over{1+\epsilon^2}}
\left[ \left(\epsilon \bU-\bu1\right)\cdot \sig1 \right] \sig1
\right)
\eqno (Defbe)
$$
where $\sig1$ is the unit vector along the axis joining the centres of the two
spheres at contact and $\chi$ is any function of velocities $\bU$ and $\bu{1}$.

On the other hand, $\Tb (1,2)$ characterizes the effect of binary collisions
between the fluid particles. It
is given by :
$$
\Tb (1,2) = \int d\sig1 (\bu{12}\cdot\sig1) \theta(\bu{12}\cdot\sig1)
\delta(\bx{12}) \left\{b_{\sig1}(1,2)-1\right\}
\eqno(DefTb)
$$
where $\bu{12}=\bu{1}-\bu{2}$, $\bx{12}=\bx{1}-\bx{2}$, 
and $b_{\sig1}$ acts as $b^{\epsilon}_{\sig1}$ in eq. \gcall{Defbe}, with 
$\epsilon=1$ (collision between equal mass particles). Note that the
dimensionless cross section of the gas, $(\sigma/\Sigma)^2=\epsilon^2$
(see eq. \gcall{Deflimit}), has been extracted from the expression
of $\Tb$ and displayed as a prefactor in eq. \gcall{EquB1}.

In eqs. \gcall{DefTme} and \gcall{DefTb}, the fluid part of the collisional transfer is not present since we consider the Grad limit : at encounters,
the positions of the particles co\"\i ncide \refto{Resibois}.

To solve the system of integro-differential equations
\gcall{EquB}-\gcall{EquB1}, we 
have to define the initial state
of the system. As in our previous work \refto{Papiers_I,Autres}, we shall assume that the fluid is initially in conditional equilibrium in the presence of  
particle B :
$$
F_{B1}^\epsilon(B,1;\tau=0)= F_{B}^\epsilon (B;\tau=0) F^{eq} (1 \vert \bX)
\eqno(Condinit)
$$
with 
$$
F^{eq} (1 \vert \bX)= \Theta (\vert \bX-\bx{1} \vert -\xi ) \phi(\bu{1})
\eqno(Equcond)
$$
where 
$$
\phi(\bu{}) = (2\pi)^{-3/2} \exp(-u^2/2)
\eqno(Max)
$$
is the Maxwell distribution and $\Theta$ is the Heavyside step function.

Before closing this section, let us recall some results concerning
the $\epsilon$ expansion of the collision operator
$\Tme (B,1)$, which will be used in the course of the analysis.  
The calculations leading to them
, though technical, are quite straightforward and have been 
presented in our previous paper (ref. \refto{Papiers_I} quoted later as I).

The B-gas collision operator can be formally expanded in powers of
$\epsilon=(m/M)^{1/2}$ as
$$
\Tme (B,1)\ =\ \Tm^{(0)} (B,1)\ +\ \epsilon\ \Tm^{(1)} (B,1)\ +
\ \epsilon^2\ \Tm^{(2)} (B,1)\ +\ ...
\eqno(DevTme)
$$
The zeroth order term $\Tm^{(0)} (B,1)$ characterizes collisions
of the gas particles with the immobile B particle,
acting as an external field. One finds
$$
\eqalign{
\Tm^{(0)} (B,1) = &\sec \int d\sig1 \ (-\bu1 \cdot \sig1) \theta (-\bu1 \cdot \sig1) \cr
&\times \Biggl\{ \delta \left( \bX - \dia \sig1 - \bx1\right) b_{\sig1}^{(0)}
(B,1) - \delta \left( \bX + \dia \sig1 - \bx1\right) \Biggr\}
}
\eqno(Tm0)
$$
where $b_{\sig1}^{(0)}$ represents the change of velocity of a gas particle, 
undergoing a specular collision with the
B particle fixed at point $\bX$ (see eq. \gcall{Defbe}, with $\epsilon=0$).

The first order correction has a much more complicated form
(see Appendix of paper I for the full expression), but we will only need
the action of $\Tm^{(1)} (B,1)$ on a state in which the
gas is in conditional equilibium \gcall{Equcond}. In this case, one
finds the following formula :
$$
\Tm^{(1)} (B,1) F(B) F^{eq} (1 \vert \bX) = F(B) \bU \cdot \dX F^{eq} (1 \vert \bX) - \Fmn (1) \cdot \left( \dU + \bU \right) F(B) F^{eq} (1 \vert \bX)
\eqno(Tm1)
$$
where the notation $\Fmpn (1)$ has been introduced to denote a dimensionless 
microscopic ``force'' 
$$
\Fmpn (1) = \sec \int d\sig1 \ \left[ 2(\bu{1}\cdot \sig1)^2 \theta 
(\mp\bu{1}\cdot \sig1) \right] \sig1\ \delta\left( \bX-\bx{1} -\dia \sig1 \right)
\eqno(DefFmp)
$$

We shall also need some formulas involving integration over the fluid 
degrees of freedom. First, one can verify that the integrated zeroth-order
term identically vanishes :
$$
\int d{\bf 1} \ \Tm^{(0)} (B,1) F_{B1} (B,1) \ = \ 0.
\eqno(Tm0integre)
$$
The first-order term is given by
$$
\eqalign{
\int d{\bf 1} \ &\Tm^{(1)} (B,1) F_{B1} (B,1) \ = \cr
&=- \int d{\bf 1}\ \Fpn (1) \cdot\dU F_{B1} (B,1)
}
\eqno(Tm1integre)
$$
with $\Fpn (1)$ defined in eq. \gcall{DefFmp}.

The general expression for the integrated second order term is more involved (see eq. (31)
of paper I for the complete formula), but when the fluid is in
conditional equilibrium, it reduces to
$$
\eqalign{
\int d{\bf 1}\ \Tm^{(2)} (B,1) &F_{B} (B) F_1^{eq}(1\vert
\bX) \cr
&= \sec  {8 \over 3} \sqrt{2 \pi} \dU \cdot \left( \dU +\bU
\right) F_B (B)
}
\eqno(Tm2integre)
$$

\vskip 0.4in{\bf 3. The Multiple Time Scale Analysis}

The system of coupled equations \gcall{EquB}, \gcall{EquB1} will be 
studied in the Brownian limit where $\epsilon \ll 1$. As in our previous
work, this will be achieved systematically by using a multiple time-scale
analysis, which leads to a uniformly convergent $\epsilon$-expansion,
avoiding secular divergences as time goes to infinity.
We just recall here the spirit of the method. In the $\epsilon \rightarrow 0$ limit,
different time-scales separate out in the system. Accordingly, we replace
the distribution functions $F_B$, $F_{B1}$ by auxiliary functions, 
$F^\epsilon_B(B;\tau_0,\tau_1,\tau_2,\dots)$, 
$F^\epsilon_{B1}(B,1;\tau_0,\tau_1,\tau_2,\dots)$ which now depend on many time arguments.
The time derivative is replaced accordingly by the operator
$$
\dta0 \ + \ \epsilon \ \dta1 \ + \ \epsilon^2 \ \dta2 \ + \ \dots
\eqno(Dtau012)
$$
The auxiliary functions are also expanded in powers of $\epsilon$ :
$$
\eqalign{
&F_B^\epsilon = F_B^{(0)} + \epsilon F_B^{(1)} + \epsilon^2 F_B^{(2)} + \dots\cr
&F_{B1}^\epsilon = F_{B1}^{(0)} + \epsilon F_{B1}^{(1)} + \epsilon^2 F_{B1}^{(2)} + \dots\cr
}
\eqno(ExpFB1)
$$
The expansions \gcall{Dtau012}, \gcall{ExpFB1} are substituted into the evolution equations \gcall{EquB} and
\gcall{EquB1}, supplemented by \gcall{Fact}, and terms of the same order in 
$\epsilon$ are identified. The determination of successive
corrections $F_B^{(k)}$, $F_{B1}^{(k)}$ ($k=0,1,2,...$) is then achieved
by combining the chosen initial condition with the requirement that the
expansion in $\epsilon$ be uniform with respect to time, which amounts to eliminating secular
divergences. The physically relevant solution of eqs. \gcall{EquB} and \gcall{EquB1}
is then obtained by restricting the originally independent multiple time variables $\tau_0$, $\tau_1$, $\tau_2$ to the ``physical line'' :
$$
\tau_0=\tau, \ \ , \tau_1=\epsilon \tau, \ \ , \tau_2=\epsilon^2 \tau,\ \dots
\eqno(15)
$$
on which the operator \gcall{Dtau012} reduces back to $\dtau$. The dependence of the 
distribution functions on the variable $\tau_{j}$ essentially defines the
dynamical evolution on the time scale $\tau\simeq \epsilon^{-j}$, $j=0,1,2,...$

Finally, we shall assume that the 
initial condition is entirely contained in the zero order
terms :
$$
\eqalign{
F^{(0)}_B(B;\tau_0=0, \tau_1=0,\tau_2=0,\dots) &= F^\epsilon_B (B;\tau=0) \cr
F^{(0)}_{B1}(B,1;\tau_0=0, \tau_1=0,\tau_2=0,\dots) &= F^\epsilon_{B1} (B,1;\tau=0) \cr
}
\eqno(Solinit)
$$

\endpage
{\bf 4. $\epsilon$-Expansion and Derivation of the Reduced Equation for the B Particle}

We first consider the zeroth-order terms in the coupled kinetic equations
\gcall{EquB}-\gcall{EquB1}. Since at this order the integrated gas-B collision operator vanishes
(see eq. \gcall{Tm0integre}), we are left with
$$
\eqalign{
&\dta0 F_B^{(0)} (B;\tau_0,\tau_1,\tau_2,\dots) =0 \cr
&\dta0 F_{B1}^{(0)} (B,1;\tau_0,\tau_1,\tau_2,\dots)=0 \cr
}
\eqno(Evol0)
$$
So both distributions,  $F_B^{(0)}$ and $F_{B1}^{(0)}$, do not depend on the time variable
$\tau_0$.

To first order in $\epsilon$, the kinetic equations reduce to
$$
\dta0 F_B^{(1)}   + \dta1 F_B^{(0)}  
=\int d{\bf 1}\ \Tm^{(1)} (B,1) F^{(0)}_B F^{(0)}_{1 \vert B} 
\eqno(Evol1 .a)
$$
$$
\eqalign{
\dta0 F_{B1}^{(1)}   + \dta1 F^{(0)}_{B1}   =
&=\int d{\bf 2}\ \Tm^{(1)} (B,2) F_B^{(0)}~ F^{(0)}_{1 \vert B} ~
F^{(0)}_{2 \vert B}\cr
}
\eqno(Evol1 .b)
$$
The action of the collision operator $\Tm^{(1)}$ is given in eq. \gcall{Tm1integre}.
Since $F_B^{(0)}$ and $F_{B1}^{(0)}$ and the r.h.s. of eqs. \gcall{Evol1}
are independent of $\tau_0$, we must impose 
$$
\eqalign{
&\dta0 F_B^{(1)} (B;\tau_0,\tau_1,\tau_2,\dots) =0 \cr
&\dta0 F_{B1}^{(1)} (B,1;\tau_0,\tau_1,\tau_2,\dots)=0 \cr
}
\eqno(Evol0bis)
$$
to eliminate secular divergences.
On the other hand, one can verify that eq. \gcall{Tm1integre} implies
$$
\int d{\bf 1}\ \Tm^{(1)} (B,1) F_B(B) F^{eq} (1 \vert \bX) = 0
\eqno(Tm1equ=0)
$$
The integrated operator $\Tm^{(1)}$
applied to a state of the system in which the
fluid is in conditional equilibrium, vanishes after integration over the 
fluid variables, 
whatever the state of particle B.
Hence, the solution of eqs. \gcall{Evol1} consistent with the initial
condition \gcall{Condinit} has the form :
$$
\eqalign{
&\dta1 F_B^{(0)} (B;\tau_1,\tau_2,\dots) =0 \cr
&F^{(0)}_{1\vert B} (1 \vert B) = F^{eq} (1 \vert \bX) \cr
}
\eqno(Sol1)
$$

The evolution of the system takes thus place on a longer time-scale,
corresponding to variable $\tau_2$. The second order terms in
the kinetic equations \gcall{EquB}-\gcall{EquB1} yield the relations
$$
\eqalign{
\dta0 & F_B^{(2)}+  \dta1 F_B^{(1)} + \dta2 F_B^{(0)}
=
\cr
&=\int d{\bf 1}\ \Tm^{(2)} (B,1) F_B^{(0)}~ F^{eq} (1 \vert \bX) 
+ \int d{\bf 1}\ \Tm^{(1)} (B,1) F_B^{(0)}~ F^{(1)}_{1\vert B}
\cr
}
\eqno(Evol2 .a)
$$
$$
\eqalign{
\dta0 &F_{B1}^{(2)} +\dta1 F_{B1}^{(1)}  + \dta2 F_B^{(0)} 
F^{eq} (1 \vert \bX) + 
\left( \bu1\cdot\dx1 - \Tm^{(0)} (B,1)\right) F_B^{(0)} F^{eq} (1 \vert \bX) \cr
=\int &d{\bf 2}\ \Tm^{(2)} (B,2) F_B^{(0)}~ F^{eq} (1 \vert \bX) ~
F^{eq} (2 \vert \bX)
+ \int d{\bf 2}\ \Tm^{(1)} (B,2) F_B^{(0)}~F^{eq} (1 \vert \bX) ~
F^{(1)}_{2\vert B}
 \cr
+\int &d{\bf 2}\ \Tb (1,2) F_B^{(0)}~ F^{eq} (1 \vert \bX) ~
F^{eq} (2 \vert \bX)
\cr
}
\eqno(Evol2 .b)
$$
In obtaining these equations, the equality 
$F^{(0)}_{1\vert B}=F^{eq} (1 \vert \bX)$ and the 
relation \gcall{Tm1equ=0} have been taken into account.

Since $F_B^{(0)}$ and $F_{B1}^{(0)}$ are independent of
$\tau_0$ and $\tau_1$, and $F_B^{(1)}$ and $F_{B1}^{(1)}$ of $\tau_0$,
we conclude from \gcall{Evol2} that in order to eliminate secular divergences one has
to impose
$$
\eqalign{
&\dta0 F_B^{(2)} (B;\tau_0,\tau_1,\tau_2,\dots)\ =\ \dta0 F_{B1}^{(2)} (B,1;\tau_0,\tau_1,\tau_2,\dots)=0 \cr
&\dta1 F_B^{(1)} (B;\tau_1,\tau_2,\dots)\ =\ \dta1 F_{B1}^{(1)} (B,1;\tau_1,\tau_2,\dots) \ =\ \dta1 F_{1\vert B}^{(1)} (1\vert B;\tau_1,\tau_2,\dots)=0 \cr
}
\eqno(Evol0ter)
$$

Now, since $F^{eq} (1 \vert \bX)$ is the conditional equilibrium of
the gas in the presence of the fixed B particle, the
following relations are satisfied :
$$
\left( \bu1\cdot\dx1 - \Tm^{(0)} (B,1)\right) F^{eq} (1 \vert \bX) = 0
\eqno(Equilibre1 .a)
$$
$$
\int d{\bf 1}\ \Tb (1,2) F^{eq} (1 \vert \bX) 
F^{eq} (2 \vert \bX) = 0
\eqno(Equilibre1 .b)
$$
In view of these equalities, both evolution equations
\gcall{Evol2 .a} and \gcall{Evol2 .b} are found to reduce to 
$$
\dta2 F_B^{(0)} = \zeta_B \dU \cdot \left( \dU +\bU
\right) F_B^{(0)} +
\int d{\bf 1}\ \Tm^{(1)} (B,1) F_B^{(0)}~ F^{(1)}_{1\vert B}
\eqno(Evol2fin)
$$
where formula \gcall{Tm2integre} for $\Tm^{(2)} (B,1)$ has been used and
we defined the Boltzmann friction coefficient $\zeta_B$ by
$$
\zeta_B=\sec  {8 \over 3} \sqrt{2 \pi}
\eqno(DefZetaB)
$$

The first term on the r.h.s of \gcall{Evol2fin} characterizes
the instantaneous ``static'' friction force induced by collisions
of the gas particles with particle B. By analogy with the hydrodynamic
expression for the friction force ($\zeta \times $ velocity), the role of the velocity
is played here by $\left( \dU +\bU
\right) F_B^{(0)} (B)$. On the other hand, the physical meaning
of the second term on the r.h.s. of \gcall{Evol2fin} can be clarified by introducing
a mean dynamic friction force $\Fmoy$ defined by
$$
\Fmoy (B;\tau_2) = \int d{\bf 1}\ \Fpn (1) F^{(1)}_{1\vert B}(1 \vert B; \tau_2)
\eqno(DefFmoy)
$$
This time-dependent force characterizes the ``dynamical'' part of the drag
on the B particle, 
induced by the correlations building up inside the gas, in response to the
motion of particle B.
Using this definition, eq. \gcall{Evol2fin} can be rewritten
$$
\dta2 F_B^{(0)} (B;\tau_2,\dots)= \dU \cdot \left\{ \zeta_B  \left( \dU +\bU
\right) F_B^{(0)} (B;\tau_2,\dots)  -\Fmoy (B;\tau_2) F_B^{(0)}(B;\tau_2,\dots) \right\}
\eqno(EvolF0)
$$
To close this equation, we have to evaluate the friction force \gcall{DefFmoy}.
To this end we must consider the third order terms in the kinetic
equations \gcall{EquB},\gcall{EquB1}. 

We first
note that the already presented reasonning concerning the secular divergences
(see eq. \gcall{Evol0ter}),
when applied to the third order equations,  yields
$$
\eqalign{
&\dta0 F_B^{(3)} (B;\tau_0,\tau_1,\tau_2,\dots)\ =\ \dta0 F_{B1}^{(3)} (B,1;\tau_0,\tau_1,\tau_2,\dots)=0 \cr
&\dta1 F_B^{(2)} (B;\tau_1,\tau_2,\dots)\ =\ \dta1 F_{B1}^{(2)} (B,1;\tau_1,\tau_2,\dots) \ =\ \dta1 F_{1\vert B}^{(2)} (1\vert B;\tau_1,\tau_2,\dots)=0 \cr
}
\eqno(Evol0qua)
$$
The evolution equations to third order in $\epsilon$ take thus the form
$$
\eqalign{
\dta2 &F_B^{(1)}  + \dta3 F_B^{(0)} + \bU \cdot \dX F_B^{(0)} = \cr
&= \int d{\bf 1}\ \Tm^{(1)} (B,1) F^{(2)}_{B1}  
+ \int d{\bf 1}\ \Tm^{(2)} (B,1) F^{(1)}_{B1}  
+ \int d{\bf 1}\ \Tm^{(3)} (B,1) F^{(0)}_{B1} \cr
}
\eqno(Evol3 .a)
$$
$$
\eqalign{
\dta2 & \biggl\{ F_B^{(0)} ~ F_{1\vert B}^{(1)}  + F^{(1)}_B~
F^{eq} (1 \vert \bX)\biggr\} + \dta3 F_B^{(0)}~ F^{eq} (1 \vert \bX)
\cr
&+ \left( \bu1\cdot\dx1 - \Tm^{(0)} (B,1)\right) F_B^{(0)}~F_{1\vert B}^{(1)} 
+\left\{ \bU \cdot \dX  - \Tm^{(1)} (B,1) \right\} F_B^{(0)}~ F^{eq} (1 \vert \bX)
\cr
&= \int d{\bf 2}\ \Tm^{(1)} (B,2) F^{(2)}_{B12}  
 + \int d{\bf 2}\ \Tm^{(2)} (B,2) F^{(1)}_{B12} 
 + \int d{\bf 2}\ \Tm^{(3)} (B,2) F^{(0)}_{B12} \cr
&+ \int d{\bf 2}\ \Tb (1,2)       F^{(1)}_{B12} \cr
}
\eqno(Evol3 .b)
$$
where the relation \gcall{Equilibre1} has already been
taken into account. The functions $F_{B1}^{(k)}=\{F_B F_{1\vert B}\}^{(k)}$ and
$F_{B12}^{(k)}=\{F_B F_{1\vert B}F_{2\vert B}\}^{(k)}$ in the
collision terms represent the $k$th order contributions $k=0,~1,~2$. For instance,
$$
\eqalign{
F^{(2)}_{B12} (B,1,2) &= \left\{F_B~F_{1\vert B}~F_{2\vert B} \right\}^{(2)} \cr
&= F^{(2)}_B ~ F^{eq} (1 \vert \bX) F^{eq} (2 \vert \bX)  \cr
&+ F^{(1)}_B ~ \biggl\{F_{1\vert B}^{(1)} ~F^{eq} (2 \vert \bX) 
+F^{eq} (1 \vert \bX) F_{2\vert B}^{(1)} \biggr\} \cr
&+ F_B^{(0)}~ \biggl\{F_{1\vert B}^{(2)} ~
F^{eq} (2 \vert \bX) + F^{eq} (1 \vert \bX)
F_{2\vert B}^{(2)}  +F_{1\vert B}^{(1)} ~F_{2\vert B}^{(1)}  \biggr\} \cr
}
\eqno(DevFB12)
$$

Our concern is now to obtain a closed equation for the conditional distribution
of the gas $F_{1\vert B}^{(1)} (1\vert B)$. To this end, we have to eliminate
the terms containing the $\tau_3$-variable, which can be achieved by
subtracting from \gcall{Evol3 .b} the first equation \gcall{Evol3 .a} 
multiplied by $F^{eq} (1 \vert \bX)$. One finds
$$
\eqalign{
\dta2 &\biggl( F_B^{(0)}~  F_{1\vert B}^{(1)} \biggr) +
\left( \bu1\cdot\dx1 - \Tm^{(0)} (B,1)\right) F_B^{(0)}~F_{1\vert B}^{(1)} =\cr
&=\ \Tm^{(1)} (B,1) F_B^{(0)}~ F^{eq} (1 \vert \bX) - F_B^{(0)}~ \bU \cdot \dX F^{eq} (1 \vert \bX)\cr
& +\int d{\bf 2}\  \Tm^{(1)} (B,2) F_B^{(0)}~ F_{1\vert B}^{(1)} ~ F_{2\vert B}^{(1)}  \cr
& + \int d{\bf 2}\ \Tm^{(2)} (B,2) F^{(0)}_B~  F_{1\vert B}^{(1)} ~
F^{eq} (2 \vert \bX)  \cr
& + \int d{\bf 2}\ \Tb (1,2)  F^{(0)}_B \left\{ F_{1\vert B}^{(1)} ~
F^{eq} (2 \vert \bX) +F^{eq} (1 \vert \bX)~ F_{2\vert B}^{(1)} 
\right\} \cr
}
\eqno(Evol3int)
$$
The last term of \gcall{Evol3int} introduces the (dimensionless) linearized Boltzmann operator
$\Lambda_B (1)$, defined as \refto{Resibois} :
$$
\Lambda_B (1) \Psi (1\vert B) = \int d{\bf 2}\  F^{eq}(1 \vert \bX) \phi(\bu2 )\ \Tb (1,2) 
\left\{ \left( {\Psi (1\vert B)\over {\phi(\bu1)}}\right) + 
\left( {\Psi (2\vert B)\over {\phi(\bu2)}}\right) \right\}
\eqno(DefLB)
$$
where $\phi(\bu{} )$ is the Maxwell distribution. Note that $\Lambda_B (1)$  acts only on the fluid
variables.

With the use of relations \gcall{Tm1}, \gcall{Tm1integre} and \gcall{Tm2integre}, the equation \gcall{Evol3int} reduces to :
$$
\eqalign{
\Biggl\{ \dta2 & + \bu1\cdot\dx1 - \Tm^{(0)} (B,1) - \Lambda_B (1) + 
\dU \cdot \biggl( \Fmoy (B) -\zeta_B \biggl(\dU + \bU \biggr) \biggr) \Biggr\}
F^{(0)}_B(B)  F_{1\vert B}^{(1)} (1 \vert B) = \cr
& = \  - \Fmn (1) \cdot \left(\dU +\bU\right) F_B^{(0)}(B)F^{eq} (1 \vert \bX) \cr
}
\eqno(EvolF01int)
$$
The notation
$$
\eqalign{
& \Lf = \bu1\cdot\dx1 - \Tm^{(0)} (B,1) - \Lambda_B (1) \cr
& \LB = \dU \cdot \left\{\Fmoy (B) -\zeta_B \left(\dU + \bU \right)\right\}
\equiv \dU \cdot \Ftot \cr
}
\eqno(Def_LfLB)
$$
and 
$$
F^{corr} (B,1;\tau_2)= F^{(0)}_B(B)  F_{1\vert B}^{(1)} (1 \vert B),
\eqno(DefFcor)
$$
permits to cast \gcall{EvolF01int} in a more transparent form
$$
\left\{ \dta2 + \Lf + \LB \right\} F^{corr} (B,1;\tau_2)
=\ 
- \Fmn (1) \cdot \left(\dU +\bU\right) F_B^{(0)}(B;\tau_2)F^{eq} (1 \vert \bX)
\eqno(EvolF01)
$$
The force $\Ftot$, appearing in the definition \gcall{Def_LfLB} of $\LB$,  can be interpreted as the total friction force acting on 
particle B.
The system of equations \gcall{EvolF0},\gcall{EvolF01} is closed and characterizes
the dynamical evolution of the distribution functions on the $\tau_2$ time-scale.

The formal solution of eq. \gcall{EvolF01} can be written as
$$
\eqalign{
F^{corr} (B,1;\tau_2)= - \int_0^{\tau_2} ds\ \exp&\left\{ {-\int_s^{\tau_2} ds^\prime\ \left( \Lf + \LB \right) (s^\prime)} \right\}
\Fmn (1) \cr
&\ \ \ \ \ \ \ \ \ \ \ \ \cdot \left(\dU +\bU\right) F_B^{(0)}(B;s) F^{eq} (1 \vert \bX) \cr
}
\eqno(SolFcorint)
$$
However, as can be verified from the definitions \gcall{Def_LfLB}
combined with relations \gcall{Tm0}, \gcall{DefLB} and \gcall{DefFmoy}, 
$\Lf$ acts only on the fluid variables, while $\LB$ acts only on the velocity
of the B particle. The two operators thus commute and the solution for $F^{corr}(B,1)$ reads
$$
\eqalign{
F^{corr} (B,1;\tau_2)= - \int_0^{\tau_2} ds\ \Fmn &(1;-(\tau_2-s)) F^{eq} (1 \vert \bX) \cr
&\cdot  \exp\left\{ {-\int_s^{\tau_2} ds^\prime\ \LB (s^\prime)}\right\} \left(\dU +\bU\right) F_B^{(0)}(B;s) \cr
}
\eqno(SolFcor)
$$
where $\Fmn (1;-\tau)$ denotes the force $\Fmn (1;0)$ on particle B propagated by the 
intrinsic fluid dynamics in the presence of the B particle, fixed at point $\bX$
(see the definition of $\Tm^{(0)}$, \gcall{Tm0}), backward in time to the instant $-\tau$.

The dynamical part of the mean friction force, $\Fmoy (B)$, defined in \gcall{DefFmoy},
can now be obtained by multiplying \gcall{SolFcor} by $\Fpn (1)$ and averaging over the
fluid variables. This yields
$$
\eqalign{
\Fmoy (B;\tau_2) F^{(0)}_B(B;\tau_2) &\equiv
\int d{\bf 1}\ \Fpn (1) F^{corr} (B,1;\tau_2) \cr
&= \int_0^{\tau_2} ds\ \Cdyn (\tau_2-s) 
\exp\left\{ {-\int_s^{\tau_2} ds^\prime\ \LB (s^\prime)}\right\} \left(\dU +\bU\right) F_B^{(0)}(B;s) \cr
}
\eqno(SolFmoy)
$$
where we have introduced $\Cdyn (\tau)$, a force-force correlation function, defined by
$$
\eqalign{
\Cdyn (\tau) &= {1\over 3} \int d {\bf 1}\ \Fmn (1;-\tau) \cdot \Fpn (1;0) 
F^{eq} (1 \vert \bX)\cr
& \equiv {1\over 3} \bra \Fmn (1;-\tau) \cdot \Fpn (1;0) \ket_\eq \cr
}
\eqno(DefCdyn)
$$
By using the notation $\bra \dots \ket_\eq$, we would like to stress the fact that 
the average is taken over the {\it equilibrium} ensemble of the fluid, 
in the presence of particle B {\it fixed} at point $\bX$.
The prefactor $1/3$ stems from the isotropy of the system around the B particle.

Moreover, eq. \gcall{SolFmoy} can be seen as a self-consistency relation for
the (time-dependent) mean force $\Fmoy$, because the propagator $\LB$
, defined in \gcall{Def_LfLB}, depends itself on the friction force $\Fmoy$. 
Substituting this expression for $\Fmoy$ in eq. \gcall{EvolF0}, one arrives at a reduced equation for the distribution of
the B particle
$$
\eqalign{
\dta2 F_B^{(0)}(B;\tau_2) = &\zeta_B \dU \cdot    \left( \dU +\bU
\right) F_B^{(0)} (B;\tau_2)  \cr
&+  \int_0^{\tau_2} ds\ \Cdyn (\tau_2-s) 
\dU \cdot \exp\left\{ {-\int_s^{\tau_2} ds^\prime\ \LB (s^\prime)}\right\} \left(\dU +\bU\right) F_B^{(0)}(B;s)\cr
}
\eqno(EvolF0final)
$$
where $\LB$ is defined in \gcall{Def_LfLB} and $\Cdyn$ in \gcall{DefCdyn}.

At this stage, we can return to the original variables $\bR$, $\bV$ and $t$,
using relations \gcall{Var}, \gcall{Tau}, \gcall{FBdim} and \gcall{Fctil}.
For the sake of simplicity, we keep the same notations for $\zeta_B$, $\Cdyn$, and for different forces and propagators involved in the evolution equations. Their full expressions are now given by
$$
\zeta_B={1 \over M} \left( {\Sigma \over 2} \right)^2 {8 \over 3}
n (2\pi m k_B T)^{1/2} 
\eqno(DefzetaBdim)
$$
$$
\Cdyn (t) = {1 \over {3Mk_BT}} \bra \Fmn (1;-t) \cdot \Fpn (1;0) \ket_\Eq
\eqno(DefCdyndim)
$$
where the formula for the microscopic forces $\Fpm$ now reads
$$
\Fmpn (1) \ =\  \left( {{ \Sigma}\over 2} \right)^2
\int d\sig1 \ 2m (\bv{1}\cdot \sig1)^2 \theta  (\mp \bv{1}\cdot \sig1) \sig1 
\ \delta \left( \bR - \left( {{ \Sigma}\over 2} \right) \sig1 -\br{1}
\right)
\eqno(DefFpm)
$$
In \gcall{DefCdyndim}, the notation $\Eq$ stands for an average over the 
equilibrium ensemble of the gas, in the presence of particle B, {\it fixed}
at $\bR$. The dynamics is characterized by the propagator
$\Lf$, given in the appendix together with $\LB$ (for the dimensionless form,
see eq. \gcall{Def_LfLB}). 

Keeping in mind that we are only interested in the dynamics of the system
on the time-scale characterized by the variable $\tau_2$, {\it i.e.}
$\tau\sim \epsilon^{-2}$
($\tau$ is the dimensionless time variable defined
in eq. \gcall{Tau}), we obtain the following equation for the
distribution function $f_B$ of the B particle
$$
\eqalign{
\dt f_B(B;t) = &\zeta_B \dV \cdot    \left( \bV + {k_BT\over M} \dV
\right) f_B (B;t)  \cr
&+  \int_0^{t} ds\ \Cdyn (t-s) 
\dV \cdot \exp\left\{ {-\int_s^{t} ds^\prime\ \LB (s^\prime)}\right\} \left( \bV + {k_BT\over M} \dV \right) f_B(B;s)\cr
}
\eqno(EvolF0dim)
$$
This should be completed by the self-consistency equation for the dynamical part $\Fmoy$ of
the friction force 
$$
\Fmoy (B;t) f_B(B;t) 
= \int_0^{t} ds\ \Cdyn (t-s) 
\exp\left\{ {-\int_s^{t} ds^\prime\ \LB (s^\prime)}\right\} \left(\bV + {k_BT\over M} \dV \right) f_B(B;s) 
\eqno(SolFmoydim)
$$

The system of equations \gcall{EvolF0dim}-\gcall{SolFmoydim} is closed and represents
the dynamical evolution of the state of the B particle on the $\tau\sim \epsilon^{-2}$ time-scale. This condition can be rewritten $\epsilon^3 \tau \ll 1$,
so that using eq. \gcall{Tau}, we conclude that eq. \gcall{EvolF0dim}
applies for times $t$ verifying
$$
t \ll {\Sigma \over {\sqrt{ {k_BT\over M} } } } \times {\rho_B \over \rho}
\sim {\Sigma \over {\sqrt{ {k_BT\over M} } } }
\eqno(TimeWindow)
$$
The spatial diffusion process has not yet started at this time scale, since
time $t$ is  very short compared to the time needed to cover B particle
radius with thermal velocity. However the fluid hydrodynamics is already
at work.
The time-scale \gcall{TimeWindow} thus characterizes the relaxation of the velocity of the B particle, while
its spatial state is not yet affected and will relax only on a longer 
time-scale.

Clearly,
the derived equation is {\it not} of a Fokker-Planck type. The main reason is that the friction force due to the bath builds up on the same time-scale as that
characterizing the gas dynamics, which leads to memory effects in the relaxation
of particle B. The friction force $\Ftot$ has thus to be constructed in a self-consistent
way and depends on the whole history of the Brownian motion.

Equation \gcall{EvolF0dim} and its systematic derivation are the main results of the present work.

\vskip 0.4in{\bf 5. Long-time Limit of the Reduced Equation}

In this section, we discuss briefly the long-time
limit of the system \gcall{EvolF0dim}-\gcall{SolFmoydim}. Note however
that we stay in the time window $\tau\sim \epsilon^{-2}$, since 
it is only there that 
these equations do represent the evolution of the system. On
longer time-scales (e.g. $\tau\sim \epsilon^{-3}$, $\tau\sim \epsilon^{-4}$),
we expect the spatial relaxation to occur through the 
Smoluchowski equation (see the note placed in the introduction) \refto{Masters}.

In the long-time limit, the B particle velocity distribution
will relax towards the stationary solution 
of \gcall{EvolF0dim}, that is towards the Maxwellian distribution.
In the final stage, we can then write
$$
\left(\bV + {k_BT\over M} \dV \right) f_B(B;t) \simeq 0
\eqno(Approx1)
$$
and the friction force $\Ftot$ will accordingly decay to zero too. Then to first
order in this quantity, we can put $\LB \simeq 0$ in the expression
\gcall{SolFmoydim} for $\Fmoy$ and obtain
$$
\Fmoy (B;t) f_B(B;t) 
\simeq \int_0^{t} ds\ \Cdyn (t-s) 
 \left(\bV + {k_BT\over M} \dV  \right) f_B(B;s)
\eqno(Approx2)
$$
The limiting form of the reduced equation for $f_B(B;t)$ 
then simplifies to :
$$
\dt f_B(B;t) \simeq \int_0^{t} ds\ 
\ \zeta (t-s) \dV \cdot \left(\bV + {k_BT\over M} \dV\right) f_B(B;s)
\eqno(F0tinf)
$$
where the time-dependent friction coefficient
$$
\zeta(t)=\zeta_B\  \delta (t) + \Cdyn (t)
\eqno(Defzetat)
$$
has been introduced. $\delta(t)$ is the Dirac distribution.

This asymptotic form for the reduced equation of the B distribution
calls for several comments. First, though much simpler than the
complete equation \gcall{EvolF0dim} (together with \gcall{SolFmoydim}),
it still exhibits a non-markovian nature. However, the non-locality
of \gcall{EvolF0dim} in velocity space -hidden in the self-consistent
definition of $\Fmoy$ in \gcall{SolFmoydim} - is now removed and the
simplified form \gcall{F0tinf} only involves the Fokker-Planck operator
$\dV \cdot \left(\bV + {k_BT\over M} \dV\right)$. In this sense, the
velocity can be seen as following a ``generalized Ornstein-Uhlenbeck process'',
still characterized by a transition probability sharply peaked around
the mean value (\ie only small jumps occur), but now keeping the memory
of its whole previous history.

In particular, whereas for the Ornstein-Uhlenbeck process the B velocity
relaxes in an exponential way \refto{Wax,VanK},
the long-time behaviour of the friction
coefficient $\zeta(t)$ modifies the nature of the thermalization process :
we expect $f_B(B;t)$ to decay
in an algebraic way towards the Maxwell distribution, since $\zeta(t)$ is
known from hydrodynamic arguments to exhibit
a $t^{-3/2}$ long time tail \refto{Hydro}. This behaviour can be 
explicitly verified on the exact solution of eq. \gcall{F0tinf},
which can be found in refs. \refto{Lebowitz69,Boon}.

\vskip 0.4in{\bf 6. Velocity Autocorrelation Function and the Stokes-Einstein Relation}

As we discussed above, the reduced equation \gcall{EvolF0dim} (together with  \gcall{SolFmoydim}) characterizes the relaxation of the velocity
of the B particle, while the spatial relaxation will occur on a longer
time scale. As we shall see in this section, this time-scale 
separation allows one to compute explicitly the velocity ACF of the
B particle. In spite of the non-local nature of the complete system
\gcall{EvolF0dim}-\gcall{SolFmoydim}, we will eventually recover the Stokes-Einstein relation
between the diffusion and the
friction coefficient \refto{Einstein}.

Let us study the dynamical evolution of the mean velocity
$\Vmoy$ of the B particle, defined as :
$$
\Vmoy (t) = \int d\bV \ \bV \ f_B (B;t)
\eqno(DefV)
$$
Since $\Vmoy (t)$ relaxes on the time scale $\tau \sim \epsilon^{-2}$,
$f_B$ will be assumed to evolve according to equation
\gcall{EvolF0dim}.
The evolution equation for $\Vmoy (t)$ can then be obtained by multiplying
\gcall{EvolF0dim} by $\bV$ and integrating over the velocity. This
yields
$$
\eqalign{
\dt &\Vmoy(t) = \int d\bV \ \zeta_B \bV \ \dV \cdot    \left( \bV + {k_BT\over M} \dV 
\right) f_B (B;t)  \cr
+  \int_0^{t} ds\ &\Cdyn (t-s) \ \int d\bV \ \bV\ 
\dV \cdot \exp\left\{{-\int_s^{t} du\ \LB (u)}\right\} \left(\bV + {k_BT\over M} \dV \right) f_B
(B;s)
\cr
}
\eqno(EvolVB1)
$$
The first term is calculated by an integration by parts, yielding
$$
\int d\bV \ \bV\ \dV \cdot    \left( \bV + {k_BT\over M} \dV 
\right) f_B (B) = - \Vmoy(t)
\eqno(Partie1)
$$
The second term needs a more careful analysis. Let us introduce the
notation
$$
\gamma (\bV;t \vert s)= \exp\left\{{-\int_s^{t} du\ \LB (u)}\right\} \left(
\bV + {k_BT\over M} \dV \right) f_B(B;s)
\eqno(Defgamma)
$$
We thus have to compute
$$
\int d\bV \ \bV \ 
\dV \cdot \gamma (\bV;t \vert s) = - \int d\bV \ \gamma (\bV;t \vert s)
\eqno(Gammaintegre)
$$
where an integration by parts has been performed.
According to \gcall{Defgamma}, $\gamma (\bV;t \vert s)$ solves the
following initial value problem
$$
\cases{
\left(\dt + \LB (t) \right) \gamma (\bV;t \vert s) = 0 & $t \ge s $ \cr
\gamma (\bV;t= s \vert s) =\left(\bV + {k_BT\over M} \dV \right) f_B(B;s) &
 $t = s $ \cr
}
\eqno(Equgamma)
$$
Integrating eq. \gcall{Equgamma} over $\bV$, one obtains
$$
\dt \int d\bV \ \gamma (\bV;t \vert s) + \int d\bV \ \LB (t)
\gamma (\bV;t \vert s) = 0
\eqno(EquIntgamma)
$$
But according to eq. (A.1) in the appendix, the action of the operator $\LB$ 
involves multiplication by the total friction force
$\Ftot$ and then derivation with respect to $\bV$. The second term in \gcall{EquIntgamma} thus
vanishes, since
$$
\int d\bV \ \LB (t)
\gamma (\bV;t \vert s)= \int d\bV \ \dV \cdot \left\{\Ftot \ 
\gamma (\bV;t \vert s)\right\} = 0
\eqno(Terme2=0)
$$
and we are left with
$$
\dt \int d\bV \ \gamma (\bV;t \vert s) =0
\eqno(Dtau2=0)
$$
We thus eventually find
$$
\eqalign{
\int d\bV \ \gamma (\bV;t \vert s) &= \int d\bV \ \gamma (\bV;t=s \vert s)
\cr
& = \int d\bV \ \left(\bV + {k_BT\over M} \dV \right) f_B(B;s) \cr
& = \Vmoy (s) \cr
}
\eqno(Partie2)
$$
Combining eqs. \gcall{EvolVB1}, \gcall{Partie1}, \gcall{Gammaintegre}
and \gcall{Partie2}, we obtain a closed equation for $\Vmoy$ of the
form
$$
\dt \Vmoy (t) = - \int_0^{t} ds\ \zeta(t-s)\ \Vmoy (s)
\eqno(EquVB)
$$
where $\zeta(t)=\zeta_B \delta(t) + \Cdyn (t)$ is the time-dependent friction
coefficient.

Introducing the Laplace transform 
$$
\tVmoy (z) = \int_0^{+\infty} dt \ \exp(-z\cdot t) \ \Vmoy (t),
\eqno(DefLap)
$$
one obtains the explicit solution for $\tVmoy (z)$ as
$$
\tVmoy (z) = { \bV_B (t=0) \over { z + \til\zeta (z)}}
\eqno(LapVB)
$$
where $\til\zeta (z)$ denotes the Laplace transform of the time-dependent
friction coefficient.

Equation \gcall{LapVB}  characterizes the relaxation of the mean velocity of
the suspended sphere in a situation in which the B particle is initially
out of equilibrium. But according to Onsager's principle of
regression of fluctuations, this equation should as well describe the relaxation
of a {\it fluctuation} of the velocity of the B particle {\it at equilibrium}
\refto{Forster}. 
This can be
explicitly verified here.

At equilibrium, the B particle velocity autocorrelation function is
defined as 
$$
\bra \bV (t) \cdot \bV (0) \ket_{eq} =
\int d{\bf B} d{\bf 1} \dots d{\bf N}\ \bV (t) \cdot \bV (0)
\rho^{eq} ({\bf B}, {\bf 1} \dots {\bf N})
\eqno(DefVV)
$$
where  $\rho^{eq} ({\bf B}, {\bf 1} \dots {\bf N})$ is the canonical
equilibrium probability density of the system, and $\bV (t)$ denotes the
velocity of the B particle propagated in time through the dynamics
of the complete system.
Formally, this relation can be rewritten 
$$
\eqalign{
\bra \bV (t) \cdot \bV (0) \ket_{eq} &=
\int d{\bf B}\ \bV (0) \cdot \int d{\bf 1} \dots d{\bf N}\ \bV (t)  
\rho^{eq} ({\bf B}, {\bf 1} \dots {\bf N}) \cr
&= \int d{\bf B}\ \bV (0) \cdot \bra \bV (t) \ket_{n.e.} F^{eq} (\bV(0)) \cr
}
\eqno(VV2)
$$
where $F^{eq} (\bV)$ is the equilibrium distribution function
of particle B alone, and $\bra \bV (t) \ket_{n.e.}$ denotes the time-dependent mean velocity of
the B particle, averaged over the fluid variables, for a given initial 
out-of-equilibrium state
of velocity $\bV (0)$.

But this mean 
velocity will verify the evolution
equation \gcall{EquVB}, derived previously by eliminating the fluid variables.
This yields for the Laplace transform of
the velocity ACF :
$$
\bra \til\bV (z) \cdot \bV (0) \ket_{eq} = { {k_BT \over M} \over { z + \til\zeta (z)}}
\eqno(SolVV)
$$
The  diffusion 
coefficient can be directly deduced from this relation, using :
$$
\eqalign{
D &\equiv \int_0^\infty dt\  \bra \bV (t) \cdot \bV (0) \ket_{eq} \cr
& = \bra \til\bV (z=0) \cdot \bV (0) \ket_{eq} \cr
}
\eqno(DefDiff)
$$
This yields
$$
\eqalign{
D& = {{k_BT \over M} \over {\til\zeta (z=0)}} \cr
& = {k_B T  \over { M\ \zeta_f } } \cr
}
\eqno(Diff)
$$
where, according to \gcall{Defzetat}, the microscopic expression for the (integrated) friction coefficient $\zeta_f$
is given by
$$
\eqalign{
\zeta_f &\equiv \int_0^{+\infty}d\tau \ \zeta( \tau) \cr
&= \zeta_B +
{1 \over {3Mk_BT}} \int_0^\infty dt \ \bra \Fmn (1;-t) 
\cdot \Fmn (1;0) \ket_\Eq \cr
}
\eqno(CoefFriction)
$$
with $\zeta_B$ given in \gcall{DefzetaBdim}, and the microscopic forces
defined in \gcall{DefFpm}.

Equation \gcall{Diff} is precisely the Stokes-Einstein relation for
the friction coefficient. 

\vskip 0.4in
{\bf 7. Friction Coefficient and the Hydrodynamic Limit}

In this section, we would like to discuss the microscopic
formula \gcall{CoefFriction} obtained for the friction coefficient. Our aim is twofold. First, we would
like to connect these results 
to the friction coefficient showing in
the response of the fluid to an imposed motion of the
B particle. And secondly, we would like to obtain, at least
in some limits, explicit expressions for the friction coefficient
as a function of the microscopic characteristics of the system
(B particle diameter, transport coefficients of the fluid, $\dots$).
This can be achieved by using the results of van Beijeren
and Dorfman (DVB), concerning the kinetic theory of hydrodynamic
flows \refto{DVB}.

In accordance with \gcall{CoefFriction}, the friction coefficient
is the sum of two terms. The first term, $\zeta_B$, characterizes
the static effect of instantaneous collisions between the B particle and the gas.
The remaining part of the friction coefficient
involves a time-integral of a force-force correlation function, reflecting  the 
dynamical correlations induced by time-displaced collisions
between the gas and the B particle. As we discussed in our
previous work (ref. \refto{Papiers_II} quoted later as II), formula 
\gcall{CoefFriction} is equivalent to the 
Kirkwood formula for smooth potentials, relating the friction coefficient to the
time-integral of the force autocorrelation function. In the case of
hard-sphere interaction however, the force is replaced by the momentum transferred
to the B particle during instantaneous
collisions and the separation into the static term and the dynamical
part then occurs (see II for further details).
Moreover, it is interesting to note that
the same microscopic expression for the friction coefficient was obtained
in the limit considered in our previous work : 
$\epsilon \rightarrow 0$, all other parameters kept constant. 
As discussed in the introduction, this limit leads to the Fokker-Planck
equation for the B particle, so that its evolution exhibits a Markovian 
nature. We thus see that the same properties of the fluid are involved in both
limits, but they enter the dynamical laws at different levels.
They are intrinsic hydrodynamic properties of the
fluid.

Let us consider the case in which the B particle has an imposed
velocity $\bU (t)$. This amounts to represent the
B particle distribution by a Dirac $\delta$-function 
in velocity space, centered on the mean velocity $\Vmoy (t)= \bU(t)$,
with no thermal fluctuations around this value. The mean friction force $\Ftot$
acting on B can be then obtained from eq. \gcall{EquVB}, determining 
evolution of the mean velocity of
the B particle 
$$
\Ftot (t) = - \int_0^{t} ds\ M\zeta(t-s)\ \bU (s),
\eqno(EquForce)
$$
which identifies $\zeta (t)$ as the time-dependent friction
coefficient in the hydrodynamic sense. We would like to stress the fact
that this result holds (as shown in section 6) although the reduced equation \gcall{EvolF0dim} is non-local
in velocity space !

Let us now consider the dynamical part of the friction coefficient, which we 
defined
in \gcall{Defzetat} and \gcall{CoefFriction}, as
$$
\eqalign{
\zeta_{dyn}(t)
&= {1 \over {3Mk_BT}} \int d{\bf 1}\ \int_0^t ds\ \Fmn (1;-s) \cdot \Fpn (1;0)
\ f^{eq} (1 \vert \bR) \cr
}
\eqno(Defzetadyn)
$$
Our aim now is to connect 
this expression to the calculations of DVB, 
who have computed the drag on a macroscopic sphere, moving with constant
velocity. The surrounding gas was assumed to obey
an extended Boltzmann equation.

To this end, we first introduce an auxiliary function $f^c (1 \vert \bR)$, defined as
$$
\eqalign{
f^c (1 \vert \bR;t) &= \int_0^t ds\ \Fmn (1;-s)  \cdot \Vchap 
\ f^{eq} (1 \vert \bR) \cr
& = \int_0^t ds\ \exp(- s\Lf ) \Fmn (1) \cdot \Vchap 
\ f^{eq} (1 \vert \bR)\cr
}
\eqno(Deffc)
$$
where $\Lf$ is the fluid propagator (see eq. (A.2) in the appendix).
$\Vchap$ is a unit vector introduced to
establish the connections with the DVB analysis.
According to \gcall{Deffc}, 
$f^c(1 \vert \bR;t)$ verifies the following differential equation :
$$
\left(\dt + \Lf \right) f^c (1 \vert \bR;t) = \Fmn (1)\ f^{eq} (1 \vert \bR)
\cdot \Vchap
\eqno(Equfc)
$$
Now if we define $\Psi(1 \vert \bR;t)$ through the relation
$$
f^c (1 \vert \bR;t) = f^{eq} (1 \vert \bR) \ \Psi(1 \vert \bR;t)
\eqno(DefPsi)
$$
the equation \gcall{Equfc} can be rewritten
$$
\left(\dt + \Lf \right) \Psi(1 \vert \bR;t) = - T^{\prime}(B,1)\ {m \over k_BT} \ \bv1 \cdot \Vchap
\eqno(EquPsi)
$$
where $\bv1$ is the velocity of a gas particle, and the operator $T^{\prime} (B,1)$ is that defined in DVB's work :
$$
T^\prime (B,1) = \left({\Sigma \over 2}\right)^2 \int d\sig1 \ 
(-\bv1\cdot \sig1) \theta (-\bv1\cdot \sig1) \delta (\bR- {\Sigma \over 2}
\sig1 -\br1) \left[b_\sig1^{(0)}(B,1)-1\right],
\eqno(DefTprime)
$$
the operator $b_\sig1^{(0)}$ changing the velocity $\bv1$ of the gas particle
to $\bv1 - 2 (\bv1\cdot \sig1) \sig1$.

Equation \gcall{EquPsi} is equivalent to the inhomogeneous
Boltzmann equation (4.4) in the DVB paper. In their work, $\Psi(1 \vert \bR;t)$ represents the dynamical correction to the distribution of the
gas, induced by recollision
events between the gas particles and the suspended sphere, which moves
with velocity $\Vchap$.
The authors have obtained approximate solutions of this equation in the
limit where the mean-free path $\ell$ is small compared to the
radius $\Sigma/2$ of the suspended sphere. This solution has been obtained
by decomposing the distribution functions into
a hydrodynamic part and a boundary layer part. We refer to their paper
for further details \refto{DVB}. Their solution may be used to compute the
dynamical part of $\zeta_{dyn}$. Indeed, 
with the aid of \gcall{Defzetadyn}, \gcall{Deffc} and \gcall{DefPsi}, the dynamical contribution can be rewritten as
$$
M k_BT\ \zeta_{dyn}(t)=
 \int d{\bf 1}\  \Fpn (1;0)
\Psi (1 \vert R;t) \ f^{eq} (1 \vert \bR) \cdot\Vchap
\eqno(ZetaPsi)
$$
where we used the isotropy of the system when introducing $\Vchap$.
In DVB, the r.h.s. of \gcall{ZetaPsi} is interpreted as
the dynamical part of the {\it drag force}
exerted by the fluid on the suspended sphere, moving with 
velocity $\Vchap$. The established connection between both approaches 
reflects the linear-response theory, and can be physically interpreted
through Onsager's principle of regression of fluctuations.

In the limit of long times, we can then use the results (6.43) and (6.44) of DVB,
to obtain to lowest order in $\ell/\Sigma$
$$
\zeta_{dyn}(t\rightarrow \infty)= {2\pi \eta \Sigma \over M}- \zeta_B
\eqno(Solzeta)
$$
where $\zeta_B$ is the Boltzmann friction coefficient, defined in
\gcall{DefzetaBdim}, and $\eta$ the shear viscosity of the gas,
estimated within the Boltzmann approximation. Collecting both
contributions $\zeta_B$ and $\zeta_{dyn}(t\rightarrow\infty)$ , we then recover,
as expected, the
Stokes law for the friction coefficient
$$
\eqalign{
\zeta_f &= \int_0^\infty dt \ \zeta(t) \cr
& = {2\pi \eta \Sigma \over M}\cr
}
\eqno(Solzetatot)
$$
In this limit, the diffusion coefficient acquires its Stokes-Einstein 
form 
$$
\eqalign{
D &\equiv {k_BT \over M\zeta_f }\cr
&={k_BT \over 2\pi \eta \Sigma } \cr
}
\eqno(DSE)
$$

\vskip 0.4in{\bf 8. Conclusions}

In this paper, we considered the Brownian motion of
a single heavy particle moving in a bath of light particles.
Our study started with a microscopic description of the system, 
in terms of the coupled dynamical
evolution equations for the distribution functions of the B particle
and of the host fluid. The latter
was assumed to evolve according to an extended Boltzmann
equation, which correctly describes the effects 
of 
collisions between the gas particles and the suspended sphere.

Our aim was to derive a reduced equation for the B particle,
by eliminating the gas degrees of freedom in the limit of small
mass ratio $\epsilon=(m/M)^{1/2}$. However, even in the 
$\epsilon \rightarrow 0$ limit, we maintained  
the mass densities of both components to be of the same order of magnitude.
The multiple time-scale analysis has been used to construct a uniform
expansion in $\epsilon$. We derived in this way a new reduced equation
\gcall{EvolF0dim}, governing the evolution of the velocity distribution of  
particle B. This equation turned out to be non-local in time and
in velocity space. The corresponding memory
terms result from building up of the friction force by the reaction of the suspending gas
to the motion of B.
However, in spite of its non-markovian character, this reduced
equation allowed to compute the velocity
autocorrelation function of the B particle \gcall{SolVV}. Moreover,
 we recovered explicitly the
Stokes-Einstein law, relating the (spatial) diffusion coefficient
to the friction coefficient \gcall{Diff}. Finally,
the derivation yielded a microscopic expression for the friction coefficient,
reducing to the Stokes law in the hydrodynamic limit, where
the mean-free path of the gas is small compared to the radius of the
suspended sphere \gcall{Solzetatot}.

Our work is not the first attempt to obtain the reduced
equation for the distribution function of the B particle
from a microscopic point of view \refto{Lebowitz, Cukier, Mercer, Deutch,
Papiers_I,Autres}. 
However, all previous approaches derive a (local)
Fokker-Planck equation governing the dynamical evolution
of the B particle by taking the $m/M \rightarrow 0$ limit
with
all other parameters kept constant. Such a limit 
implies the asymptotically vanishing mass density ratio, while
in the present work, we maintained this ratio at a constant value.
This crucial difference is the source of the non-markovian character
of the reduced
equation for the velocity distribution of the B particle, since 
the velocity decays then on a hydrodynamic time-scale of the
suspending bath.

The results derived here suggest some directions for future work.
First, the extended Boltzmann equation was taken as a starting
point, which can be expected to be valid in the Grad limit. This conjecture could
perhaps be verified explicitly by starting from the 
BBGKY hierarchy for hard-spheres \refto{Resibois}, and applying directly
to it the present limit defined in \gcall{Deflimit}. In particular,
we could in principle compute the $\epsilon$ expansion of correlations induced in the bath by the
presence of particle B.
It would be then possible to check whether the results obtained in this work
become indeed {\it exact} in the limit \gcall{Deflimit}. 

On the other hand, it would be interesting to generalize the
analysis to an arbitrary number of suspended particles, in order
to obtain the generalized reduced equation for the $N$ 
particle suspension. In this case, we expect the hydrodynamic interactions
between the different suspended particles to be non-local in time
too.

The reduced equation we obtained here characterizes
the relaxation of the velocity distribution of B, while 
the spatial distribution evolves on a much longer time-scale. Because
of this time-scale separation, one then expects the spatial
distribution to relax according to the local Smoluchowski
equation. In principle, one could recover this equation from the same microscopic
approach by pursuing the expansion up to the corresponding time-scale,
$t\sim \Sigma^2/D$, which corresponds to the $\tau_4$ variable. Because
of the complexity of the analysis, we have performed up to now 
the corresponding 
calculations only for the case of an ideal (collisionless) suspending gas.
Even in this case, in spite of the absence of
hydrodynamic modes in the bath,
memory effects are still present and they show once two or more
Brownian particles are present. The work along these lines is being 
carried presently.

{{\bf Aknowledgements}

The authors thank Jean-Pierre Hansen for stimulating discussions
and comments on this work. J.P. aknowledges support of the P.A.S.T.
exchange program, and a partial support by KBN (Committee for Scientific
Research, Poland).}

\endpage

\head{Appendix}

\taghead{A.}
In this appendix, we give the full expressions for the 
propagators $\LB$ and $\Lf$. 

With proper dimensions, $\LB $ can be written
$$
\LB (t) = \dV \cdot \left( \zeta_B \left(\bV + {k_BT\over M} \dV \right) -
\Fmoy (B;t) \right)
\eqno(DefLBdim)
$$
where the mean force $\Fmoy (B;t)$ is defined self-consistently by
eqn. \gcall{SolFmoydim}. On the other hand, the fluid propagator $\Lf$ reads :
$$
\Lf (1) = \bv1 \cdot \dr1 - \Tm^{(0)} (B,1) - \Lambda_B (1)
\eqno(DefLfdim)
$$
where $\Lambda_B (1)$ is the linearized Boltzmann operator, 
and $\Tm^{(0)} (B,1)$ characterizes the effect of collisions between and
the gas and the B particle, fixed at point $\bR$. The
expression of $\Lambda_B (1)$ is given by \refto{Resibois} :
$$
\eqalign{
\Lambda_B (1) \Psi (1 \vert B) = f^{eq}(1 \vert \bR)\  &\int d{\bf 2}\ \int d\sig1\ \left[(\bv1-\bv2)\cdot \sig1\right]
\theta \left[(\bv1-\bv2)\cdot \sig1\right]\ \delta (\br1-\br2) \phi(\bv2) \cr
& \times \phi(\bv2) (b_\sig1 (1,2) -1) \ \left\{ \left({\Psi (1 \vert B)\over {\phi(\bv1)}}\right)
+\left({\Psi (2 \vert B)\over {\phi(\bv2)}}\right)\right\} \cr
} 
\eqno(DefLambdadim) 
$$
where $\Psi (1\vert B)$ is a function of the fluid variables 
$1 \equiv\{\br1,\bv1\}$, and $\phi(\bv{})$ is the Maxwell distribution 
$$
\phi(\bv{}) = \left({m\over {2\pi k_BT}}\right)^{3/2} \exp\left(
-{ mv^2 \over {2k_BT} }\right)
\eqno(Defmaxdim)
$$
The operator $b_\sig1 (1,2)$ acting on a function $\chi (\bv{1},\bv{2})$ replaces
the velocities $(\bv{1},\bv{2})$ by their post-collisional
values
$$
\left[b_{\sig1} (1,2) \chi \right] \left( \bv{1},\bv{2} \right)
= \chi \left( \bv{1} - \left((\bv{1}-\bv{2})\cdot \sig1\right) \sig1,
\bv{2} +  \left( (\bv{1}-\bv{2})\cdot \sig1\right) \sig1 \right)
\eqno(12)
$$

Finally, $\Tm^{(0)} (B,1)$ can be written as
$$
\eqalign {
\Tm^{(0)} (B,1) = &\left({{\Sigma} \over 2}\right)^2
\int d\sig1 \ \left[ - \bv1 \cdot \sig1 \right]
\theta \left[  - \bv1\cdot \sig1 \right] \cr
&\times\left\{ \delta\left(\bR-{{\Sigma} \over 2}\sig1 
-\br1\right) b^{(0)}_{\sig1} (B,1) - 
\delta\left(\bR+{{\Sigma} \over 2}\sig1 
-\br1\right) \right\}.
}
\eqno(DefTm0dim)
$$
where $b^{(0)}_{\sig1} (B,1)$ changes the velocity $\bv1$ of the gas particle
into  $\bv1^\prime= \bv1 - 2\left[ \bv1 \cdot \sig1
\right]\ \sig1$.

\endpage

\refis{Bedeaux} D. Bedeaux and P. Mazur, {\it Physica A} {\bf 76}, 247 (1974).

\refis{VanK} N.G. van Kampen, {\it Stochastic Processes in Physics
and Chemistry}, North-Holland, Amsterdam (1990).

\refis{Wax} N. Wax (editor), {\it Selected papers on Noise
and Stochastic Processes}, Dover Publications, New-York (1954).

\refis{Einstein} A. Einstein, {\it Investigations on the Theory of the
Brownian Movement}, Dover Publications, New-York (1956).

\refis{Lebowitz} J.L. Lebowitz and E. Rubin, {\it Phys. Rev.} {\bf 131},
2381 (1963);
P. R\'esibois and H.T. Davis, {\it Physica} {\bf 30}, 1077 (1964);
J.L. Lebowitz and P. R\'esibois, {\it Phys. Rev.} {\bf 139}, 1101 (1965).

\refis{Cukier} R.I. Cukier and J.M. Deutch, {\it Phys. Rev.} {\bf 177}, 240 (1969).

\refis{Mercer} J. Mercer and T. Keyes, {\it J. Stat. Phys.} {\bf 32}, 35 (1983).

\refis{Deutch} J.M. Deutch and I. Oppenheim, {\it J. Chem. Phys.} {\bf 54},
3541 (1971); T.J. Murphy and J.L. Aguirre, {\it J. Chem. Phys.} {\bf 57}, 2098
(1972).

\refis{Papiers_I} L. Bocquet, J. Piasecki, and J.P. Hansen,
 {\it J. Stat. Phys.} {\bf 76}, 505 (1994).

\refis{Autres}  L. Bocquet,
J.P. Hansen and J. Piasecki, {\it Il Nuevo Cimento} {\bf 16}, 981 (1994);
J. Piasecki, L. Bocquet and J.P. Hansen, {\it Physica A} {\bf 218}, 125 
(1995).

\refis{Papiers_II} L. Bocquet, J.P. Hansen and
J. Piasecki, {\it J. Stat. Phys.} {\bf 76}, 527 (1994).

\refis{Hauge} E.H. Hauge and A. Martin-L\"of, {\it J. Stat. Phys.} 
{\bf 7}, 259 (1973).

\refis{Hinch} E.J. Hinch, {\it J. Fluid. Mech.} 
{\bf 72}, 499 (1975).

\refis{Masters} A.J. Masters, {\it Mol. Phys.} 
{\bf 57}, 303 (1986).

\refis{Roux} J.N. Roux, {\it Physica A} {\bf 188}, 526 (1992).

\refis{Lorentz} H.A. Lorentz, {\it Lessen over Theoretische Natuurkunde.
Vol. V. Kinetische Problemen}, E.J. Brill, Leiden (1921).

\refis{DVB}  H. van Beijeren and J.R. Dorfman, {\it J. Stat. Phys.} {\bf 23},
335 (1980).

\refis{Resibois} P. R\'esibois and M. De Leener, {\it Classical Kinetic
 Theory of Fluids}, Wiley, New-York (1977).

\refis{Spohn} H. Spohn, {\it Rev. Mod. Phys.} {\bf 53}, 569 (1980).

\refis{Forster} D. Forster, {\it Hydrodynamic Fluctuations, Broken Symmetry
and Correlation Functions}, Benjamin, Reading, Mass. (1975).

\refis{Hydro} B.J. Alder and T.E. Wainwright, {\it Phys.Rev. A} {\bf 1},
18 (1970); R. Zwanzig and M. Bixon {\it Phys.Rev. A} {\bf 2},
2005 (1970).

\refis{Lebowitz69} J.L. Lebowitz, J.K. Percus and J. Sykes, {\it Phys.Rev.}
{\bf 188}, 487 (1969).

\refis{Boon} J.P. Boon and S. Yip, {\it Molecular Hydrodynamics},
Dover Publications, New-York (1991).

\references
\endreferences

\endpage
\bye